# Examining Older Adults' Information Exposure, Wellbeing, and Adherence to Protective Measures During the COVID-19 Pandemic

**Nurul Suhaimi, MSc[1], Nutchanon Yongsatianchot, MSc[1], Yixuan Zhang, MSc[2], Anisa Amiji, BSc[1], Shivani A. Patel, PhD[3], Stacy Marsella, PhD[1], Miso Kim, PhD[1], Jacqueline Griffin, PhD[1], Andrea G. Parker, PhD[2]**

[1]Northeastern University, Boston, MA, USA; [2]Georgia Institute of Technology, Atlanta, GA, USA; [3]Emory University, Atlanta, GA, USA

**Abstract**

*Older adults are at greater risk of experiencing negative physical and psychological impacts of the novel coronavirus 2019 (COVID-19) pandemic. Our ongoing study is assessing COVID-19 information exposure in adults aged 55 and above compared to other age groups living in Massachusetts and Georgia. This work investigates the potential association between information exposure and wellbeing as well as adherence to COVID-19 protective measures. Our initial results show that older adults received information related to COVID-19 less frequently than the middle-aged group, yet they feel more content and less stressed than the other age groups. Further analysis to identify other potential confounding variables is addressed.*

**Introduction**

78% of US adults aged 55 and above (older adults) are reported to have one or more severe underlying medical conditions[1]. Adults in this category are also more likely to experience social isolation, loneliness, and psychological distress that are significant to negative health outcomes[2,3]. The rapid spread of COVID-19 in the country has put older adults in jeopardy, not only by the virus transmission but also by the COVID-19 information they receive. Compared with younger adults, older adults are more likely to trust the information they receive during crisis situations[4,5]. Studies related to information-seeking behavior and its relationship with psychological wellbeing during the COVID-19 pandemic are gaining attention[6,7], but the focus on older adults is still lacking. In this ongoing study, we seek to examine what information about COVID-19 adults in different age groups have been exposed to, and the relationship between the effects of consuming this information on their psychological wellbeing and adherence to protective measures guidelines. Our research questions have centered on examining how older adults' exposure to COVID-19 information differs from other age groups and how these differences affect their feelings of being present in their daily lives, empowered, stressed, and worried, along with expressions of stigmatization towards individuals of Chinese descent. Further, we assessed how the differences in their feelings might be associated with the adherence to protective measures among older adults and other age groups.

**Methodology**

We conducted an online survey of 577 adults, aged 18 and above in Georgia and Massachusetts in July 2020, asking about the information participants received about COVID-19 and their wellbeing, experiences, and change in behavior during the pandemic. Subjects were recruited through Qualtrics between 7/17/2020 and 8/17/2020. The study has been approved by the Institutional Review Board at our institution. In this paper, we focus on three categories of results. First, information exposure is measured by the frequency of receiving five types of COVID-19 information (1 = never, 5 = every day). These categories are: 1) spread and seriousness, 2) causes, symptoms, and transmission, 3) guidance on reducing risks, 4) available support and services, and 5) the impact of the virus. Second, the effect of the information received is measured through 1) positive feelings – at peace, empowered, sense of control, present in daily life, 2) negative feelings – stressed, worried, anxious, or depressed, and 3) expressions of stigma, namely, the perception that the Chinese are to blame for causing the pandemic (1 = strongly disagree, 5 = strongly agree). Lastly, we measure their adherence to recommended COVID-19 prevention behaviors of using a mask, social distancing, and not attending social gatherings (1 = does not describe me, 5 = describes me exceptionally well). Cronbach's alpha is used to estimate the internal reliability coefficients of the survey. The mean scores from each category are calculated, and descriptive statistics (mean and standard deviation) for each age group (18-24, 25-34, 35-44, 45-54, 55+) are derived. Multiple linear regression modeling is performed to assess the significance of age for each measured category.

**Results**

Adults 55 and above scored significantly lower than adults in the 45-54 age group in the frequency of receiving information related to the spread and seriousness of COVID-19 (3.26 +/- 1.05 vs 3.41 +/- 1.07, p<0.001), causes, symptoms, and transmission of the virus (3.28 +/- 0.93 vs 3.38 +/- 1.13, p=0.006), and guidance on reducing risks (3.79 +/- 0.92 vs 3.88 +/- 0.87, p<0.001). Information related to available support and services during COVID-19 and the impact of COVID-19 was received more frequently by adults in the 35-44 age group (2.52 +/- 1.19, 3.37 +/- 0.98, p=0.04). Younger adults (age 18-24) had the lowest mean scores in all five information type categories.

In terms of the effect of the information they received, older adults rated the highest in feeling at peace, sense of control, empowered, and present (3.52 +/- 0.80), while younger adults (age 18-24) reported the lowest for the same feelings (2.74 +/- 0.89, p<0.001). Feeling stressed, worried, and anxious are more prevalent in adults 25-34 years old (3.92 +/- 0.98), while adults 35-44 years old reported these negative feelings the least (3.49 +/- 1.26, p=0.015). The perceptions that the Chinese were to blame for causing the pandemic had the lowest mean scores when compared to other feelings (2.52 +/- 1.334), with older adults scoring the highest (2.93 +/- 1.30, p<0.001) among other age groups. Adherence towards protective measures was rated the highest by older adults (4.39 +/- 0.97), but lowest among adults age 18-24 (3.93 +/- 0.98, p=0.02).

**Discussion**

In this study of 577 adults between July 17 to August 18, 2020, we found that there are significant differences between age groups in terms of outcomes of the frequency of the information received, the implications of the information received – feeling at present and in control, feeling stressed, and stigmas towards the Chinese, as well as the adherence towards protective measures. Among all the types of information studied, guidance on reducing risks was received most frequently, and information related to available support and services were received the least. These results may be impacted by the study's timing, as both Massachusetts and Georgia were in the middle of reopening phases (Phase 2 and 3); thus, information related to reducing virus transmission was being widely communicated to the public. On the other hand, information regarding available support and services, while publicly available, required active searches, as such information may only be of interest to specific groups.

Our results revealed that older adults were more content and less stressed than other age groups during this critical time – the opposite of findings from studies discussing older adults' wellbeing[3, 8-10]. We present two possible reasons for these findings. First, our questions regarding respondents' wellbeing are centered towards the information they received, while other studies considered social distancing as a factor of anxiety and depression among older adults[9, 10]. Second, our questions regarding how information about COVID-19 makes respondents feel may be impacted by other confounding factors, such as the platform they used to receive the information (e.g., TV, radio, mobile phone) or the type of media outlet producing the information (e.g., The New York Times, Fox News, etc.). Further analysis of these factors is warranted.

Lastly, our findings related to respondents' perceptions towards the Chinese based on the information they received, while low, are troubling. The media outlets, individuals, and organizations that our respondents trust to deliver COVID-19 information may shape the stigmatizing viewpoint that Chinese individuals are to blame for the pandemic. The driving factor behind our finding that older adults in our sample held firmer beliefs in such stigma, may be related to prior work showing that older adults are associated with a higher level of trust in general than other age groups[11]. Further research is needed to investigate these trends.

**Conclusion**

We presented our initial results regarding the differences between older adults and other age groups in COVID-19 information exposure, feelings, and adherence towards protective measures. The study still warrants a more in-depth analysis of the factors contributing to older adults' perceptions and wellbeing during the COVID-19 pandemic. In our ongoing work, we are investigating other potential factors that may shape information exposure and its impacts, such as race, education, income level, type of health insurance, and household living situation, which may provide insights for future research. Further, we aim to assess any changes in respondents' experiences and attitudes over time and correlate the changes with public health guidelines or key events during the pandemic.